\documentclass{aa}
\usepackage{pstricks}
\usepackage[]{natbib,amsmath,amssymb}
\usepackage{graphicx}
\bibpunct{(}{)}{,}{a}{}{,}

\def\RXJ{\object{RX~J1347.5$-$1145}}
\def\arcsecf{\!\!^{\prime\prime}}
\def\arcminf{\!\!^{\prime}}

\begin{document}
\title{Strong and weak lensing united II: the cluster mass distribution
  of the most X-ray luminous cluster \RXJ \thanks{Based on
  observations collected at the European Southern Observatory, Chile
  (ESO programme 67.A-0427(A-C)).}}
\titlerunning{The cluster mass distribution
  of the most X-ray luminous cluster \RXJ}
\author{M. Brada\v{c} \inst{1,2,3} \and T. Erben \inst{1} \and P. Schneider \inst{1} \and
 H. Hildebrandt\inst{1}  \and M. Lombardi
  \inst{1,4,5} \and M. Schirmer \inst{1,6} \and
  J.-M. Miralles\inst{1,4} \and  D. Clowe\inst{7}
  \and S. Schindler \inst{8}
}
\offprints{Maru\v{s}a Brada\v{c}}
\mail{marusa@astro.uni-bonn.de}
\institute{Institut f\"ur Astrophysik und Extraterrestrische
  Forschung, Auf dem H\"ugel 71, D-53121 Bonn, Germany
  \and Max-Planck-Institut f\"ur Radioastronomie, Auf dem
  H\"ugel 69, D-53121 Bonn, Germany \and KIPAC, Stanford University, 2575 Sand Hill Road, Menlo Park, CA 94025, USA
  \and European Southern Observatory, Karl-Schwarzschild-Str. 2,
  D-85748 Garching bei M\"unchen, Germany 
\and Universit\`{a} degli Studi di Milano,  v. Celoria 16, I-20133
Milano, Italy  
\and Isaac Newton Group of Telescopes, Calle Alvarez Abreu 68,
38700 Santa Cruz de La Palma, Tenerife, Spain 
 \and
Steward Observatory, University of Arizona, 933 N Cherry Ave., Tucson,
AZ 85721, USA
\and
Institute for Astrophysics, University of Innsbruck,
Technikerstr. 25, A-6020 Innsbruck, Austria
}
\date{Received 22 October 2004 / Accepted 3 March 2005}
%
%

\abstract{ We have shown that the cluster-mass reconstruction method
which combines strong and weak gravitational lensing data, developed
in the first paper in the series, successfully reconstructs the 
mass distribution of a simulated cluster. In this paper we apply the method to the
ground-based high-quality multi-colour data of \RXJ, the most X-ray
luminous cluster to date. A new analysis of the cluster core on
 very deep, multi-colour data
  analysis of  VLT/FORS data
reveals many more arc candidates than previously known
for this cluster. The combined strong and weak lensing reconstruction
confirms that the cluster is indeed very massive. If the redshift and
identification of the multiple-image system as well as the redshift
estimates of the source galaxies used for weak lensing are correct, we
determine the enclosed cluster mass in a cylinder to $M(<360\: {\rm
h}^{-1}\mbox{kpc})= (1.2 \pm 0.3) \times 10^{15} M_{\odot}$. In
addition the reconstructed mass distribution follows the distribution
found with independent methods (X-ray measurements, SZ).  With
higher resolution (e.g. HST imaging data) more reliable multiple
imaging information can be obtained and the reconstruction can be
improved to accuracies greater than what is currently possible with
weak and strong lensing techniques.  }

\maketitle
%
%

\def\diff{\mathrm{d}}
\def\ngx{N_{\mathrm{x}}}
\def\ngy{N_{\mathrm{y}}}
\def\eck#1{\left\lbrack #1 \right\rbrack}
\def\eckk#1{\bigl[ #1 \bigr]}
\def\rund#1{\left( #1 \right)}
\def\abs#1{\left\vert #1 \right\vert}
\def\wave#1{\left\lbrace #1 \right\rbrace}
\def\ave#1{\left\langle #1 \right\rangle}

%
%
\section{Introduction}
\label{sec:introduction}

Clusters of galaxies have been a focus of a very intense ongoing
research. Especially important for many cosmological applications is a
good determination of their mass. One way to obtain their masses is to
use the gravitational lensing information, both from multiple image
systems (strong lensing) as well as from distortions of background
sources (weak lensing). Many weak and strong lensing cluster mass
reconstructions have been successfully performed in the past (see e.g.
\citealt{clowe01, clowe02, gavazzi04} for examples of weak lensing and e.g.
\citealt{kneib03,smith04} for a combination of weak and strong
lensing). While weak lensing mass reconstructions have an advantage in
constraining the mass at much larger radii than strong lensing, one of
main limitations for both strong and weak lensing 
is the problem of the mass-sheet degeneracy
(i.e. the mass profile of the cluster can only be determined up to a
constant).  In the absence of redshift information from individual
sources and the lens, one can break this degeneracy only by making assumptions
about the underlying potential. Different assumptions, however, can
lead to discrepant results on the cluster mass.  In this work we
therefore use individual redshifts of background sources to overcome
this problem.  As shown in \citet{bradac03b}, by using these and by
extending the reconstruction to the inner parts of the cluster we are
effectively able to break this degeneracy.

\defcitealias{bradac04a}{Paper I} This is the second of the series of
papers in which we develop and test a cluster mass reconstruction
technique that combines strong and weak lensing information. In
\citet{bradac04a} (hereafter \citetalias{bradac04a}) we describe the
method in which we extend the weak lensing formalism to the inner
parts of the cluster, use redshift information of the background
sources and combine these with the constraints from multiply imaged
systems. Using simulated data we have shown that the method is
successful in reconstructing the mass distribution of a cluster, and
yields an excellent agreement between the input and reconstructed mass
also on scales within and beyond the Einstein radius.

Encouraged by the success of our method, we apply it to the weak
and strong lensing data for the redshift $0.451$ cluster {\RXJ}
\citep{schindler95}, the most X-ray luminous cluster known to date. Due to
its record holding, this cluster has been a subject of many studies in
X-ray \citep{schindler95, schindler97,allen02,ettori04,gitti04} and
optical \citep{fischer97,sahu98,cohen02,ravindranath02}. It has also
been detected through the Sunyaev-Zel'dovich effect
\citep{pointecouteau01,komatsu01,kitayama04}. Yet the mass
determinations based on X-ray properties, SZ effect, velocity
dispersion measurement, strong and weak lensing have all yielded
discrepant results (see \citealp{cohen02} for a summary).

For the purpose of mass reconstruction we use VLT/FORS data on a field
of $3.8\times3.8\mbox{ arcmin}^2$ in U, B, V, R, and I bands.  We also
use Ks-band data from VLT/ISAAC to obtain more reliable photometric
redshift estimates.  The shape measurements for the weak lensing
reconstruction is performed on two FORS bands, R and I. The strong
lensing properties of this cluster are analysed. From previous data
sets five arc candidates were reported \citep{schindler95,sahu98};
using the new multi-colour data we conclude that only two possibly
belong to the same multiple image system. Furthermore, we searched for
additional images belonging to this system and identified a third
possible member. Several new arc candidates were found as well and are
presented in this work.  Particularly interesting is a very red arc
candidate with two components, located at a distance of $1\mbox{
arcmin}$ from the brightest cluster galaxy (BCG). In addition, we
detect further elongated structures, some of them have been previously
indicated by \citet{lenzen04} who developed and use an automated arc
searching routine.

This paper is organised as follows. In Sect.~\ref{sc:obs} we describe
the observations and give a brief outline of the data reduction
process. In Sect.~\ref{sc:RXJ_strong} we describe how we search for
multiply imaged systems. In Sect.~\ref{sc:RXJ_rec} we give the results
of a combined strong and weak lensing reconstruction and the cluster
luminosity measurements. We conclude in Sect.~\ref{sc:RXJ_concl}.

\section{Observations and data reduction process}
\label{sc:obs}
The optical VLT data for the current project were obtained with ESO
proposal 67.A-0427(A-C) (P.I. S. Schindler).  The data was taken with
FORS1 in the high-resolution mode (pixel scale $0.\arcsecf09$; total
field of view $\approx 3.\arcminf 2\times 3.\arcminf 2$) in service
mode between April and September 2001. UBVRI Bessel filters were used
in sub-arcsecond seeing conditions (see Table~\ref{tab:data} for a
summary of data properties). This allows us to estimate
photometric redshifts for all galaxies and to support our mass and
light analysis by a careful separation of foreground and background
galaxies and cluster members (see below). Our primary band for the
weak lensing analysis (the I) was taken in the 1-port read-out
mode. Thus we avoid potential problems for object shape measurements
due to varying noise properties in the central parts of the images.
For the other 4 bands, primarily used for object photometry, the
4-port read-out mode was used. The data in each band consist of at
least 20 individual exposures and were obtained with a dither pattern
of $30.\arcsecf 0$ in RA and DEC in order to obtain clean coadded
images of highest quality.

The data reduction was carried out with a pre-release version of
\textit{THELI}, a pipeline developed specifically for the processing
of optical single- and multi-chip cameras (see \citealp{schirmer03,erben05});
here we only outline our astrometric calibration which is essential
for weak lensing studies. First, we match object positions from I-band
data with those from the USNO-A2 astrometric catalogue \citep{usno},
which fixes the position of the individual exposures with respect to
absolute sky coordinates and thus corresponds to a zero-order
astrometric solution ("shift only"). Next, we used Mario Radovich's
Astrometrix (see \citealp{mccracken03} and {\tt
http://www.na.astro.it/\~{}radovich/WIFIX/}) to fit image distortions
by a two-dimensional, third-order polynomial. Hereby, the distances of
the objects with coordinates in USNO-A2 catalogue and of the overlap
sources in different images are minimised simultaneously in the
$\chi^2$ sense. We end up with rms residuals of $\approx 0.\arcsecf
25$ for the USNO-A2 standard sources and $\approx 0.\arcsecf 01$ for
the overlap objects.  Afterwards, we extract high $S/N$ objects from
the coadded I-band image which are used as astrometric standard
sources (instead of USNO-A2) for the other bands. In all bands we
achieve formally an internal astrometric accuracy of $\approx
0.\arcsecf 01-0.\arcsecf 015$ for the overlap sources. Most of the
observations were done during photometric nights.  Photometric
zeropoints were deduced from the images of standard stars obtained as
part of the standard calibration plan of the FORS1 instrument and
reduced in the same way as the science data. The obtained zeropoints
are in good agreement with the general trend analysis of the FORS1
zeropoints.  From non-photometric nights we only include images with a
maximum absorption of 0.1 mag in the coaddition process which is
performed with drizzle \citep{drizzle}.

In addition, we retrieved Ks VLT-ISAAC data (pixel scale $0.\arcsecf
1484$; field of view $\approx 2.\arcminf 5\times 2.\arcminf 5$) from
the ESO science archive (proposal ID 67.A-0095(B)). The data was
processed with the {\sl eclipse} package (see \citealp{devillard97}).

 We create the catalogue of objects using SExtractor
\citep{sextractor} in dual-image mode. The I-band image is used for
detections and the images of the other bands are only used to measure
the corresponding magnitudes.  An object is considered detected if
five adjacent pixels had a flux that exceeded the local sky noise
level by a factor of three. All magnitudes quoted in this paper are in
the Vega system. The photometric redshifts (using isophotal magnitudes
for cluster members and aperture magnitudes for background sources;
see below) of the objects were obtained using the HyperZ package
\citep{bolzonella00}.
\begin{table}
\caption{Properties of the data used in this work. The $5\sigma$
  limiting magnitudes were determined with SExtractor using an
  aperture of $2^{\prime\prime}$.}
\begin{center}
\begin{tabular}{c c c c}
\noalign{\smallskip}
\hline
\noalign{\smallskip}
\hline
\noalign{\smallskip}
Filter & Exposure time (s) & Seeing ($''$) & Limiting mag. \\
\noalign{\smallskip}
\hline
\noalign{\smallskip}
$U$  & 11310          & 0.97 & 25.4 \\
$B$  & 4800           & 0.67 & 26.9 \\
$V$  & 4500           & 0.62 & 26.5 \\
$R$  & 6000           & 0.67 & 26.3 \\
$I$  & 6750           & 0.57 & 25.6 \\
$Ks$ & $\approx 7200$ & 0.73 & 21.4 \\ 
\noalign{\smallskip}
\hline
\noalign{\smallskip}
\end{tabular}
\end{center}
\label{tab:data}
\end{table}
\subsection{Cluster member catalog}
\label{sc:catclusters}
\begin{figure}[hbt]
\centerline{\includegraphics[width=8cm]{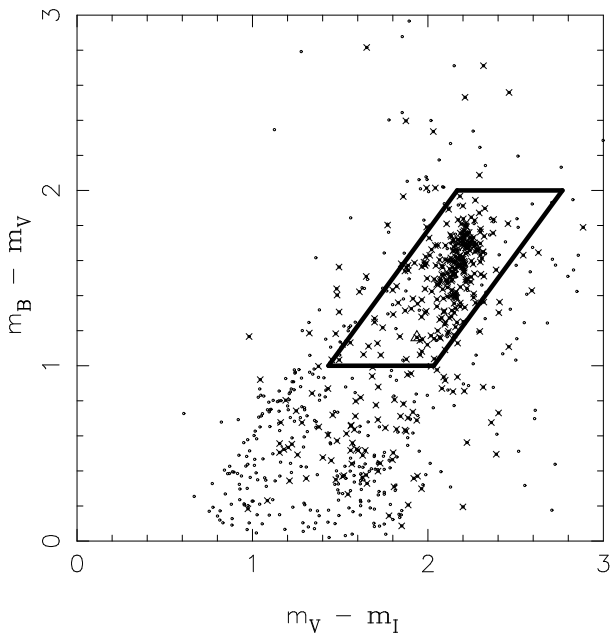}}
\caption{The $m_{\rm B} - m_{\rm V}$ vs. $m_{\rm V} - m_{\rm I}$ colours for
the galaxies in our field. Cluster members are selected to lie inside
the polygon. BCG colours are given as a triangle. In addition we
plot as crosses all the galaxies which have a photometric redshift estimate
$0.4<z_{\rm phot}<0.5$.}
\label{fig:colcol}
\end{figure}
\begin{figure}[ht]
\begin{minipage}{8cm}
\begin{center}
\includegraphics[width = 0.9\textwidth]{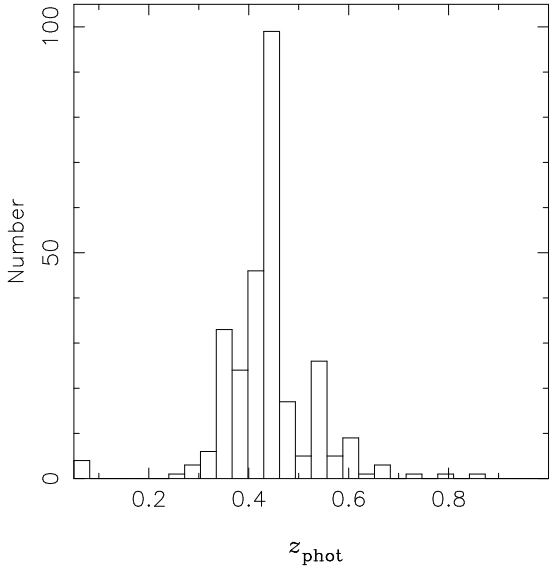}
\end{center}
\end{minipage}
\caption{The photometric 
redshift distributions of the cluster members selected as
described in Sect.~\ref{sc:catclusters}.}  
\label{fig:zphot_cl}
\end{figure}

For cluster members, due to their brightness and different sizes, we
argue that it is best to use isophotal magnitudes to obtain accurate
colour estimates. In Fig.~\ref{fig:colcol} we plot the $m_{\rm B} -
m_{\rm V}$ vs. $m_{\rm V} - m_{\rm I}$ colours for the galaxies in our
field. To determine the colour cuts for the cluster member selection
we first inspect the galaxies having I-band magnitude up to three
magnitudes fainter than the BCG and with a distance to the BCG smaller
than $1^{\prime}$. These are preferentially the cluster members and
form a group around $m_{\rm V} - m_{\rm I}\sim 2$ and $m_{\rm B} -
m_{\rm V} \sim 1.8$.  Using this information we determine the
following selection criteria for the cluster members
\begin{eqnarray}
\nonumber
0.7\rund{m_{\rm B} - m_{\rm V}+1} < &m_{\rm V} - m_{\rm I}& < 0.7\rund{m_{\rm
  B} -  m_{\rm V} + 1.9}\\
1 < &m_{\rm B} - m_{\rm V}& < 2 \;,   
\label{eq:colours}
\end{eqnarray}
and we also cut out all the objects having magnitudes brighter than
the BCG. In Fig.~\ref{fig:colcol} we plot colours for all galaxies in
our field, the BCG colours (slightly bluer than most other members)
are given as a triangle and the polygon indicates the selection
criteria we use.  In addition, to avoid biases toward red cluster
members, we add to the catalogue the blue galaxies with photometric
redshifts $0.4<z_{\rm phot}<0.5$ (denoted as crosses in
Fig.~\ref{fig:colcol}). The final photometric redshift distribution of
the cluster members is given in Fig.~\ref{fig:zphot_cl}. The
completeness of our catalogue is discussed in Sect.\ref{sc:light}.

An alternative approach would be to select the cluster members purely
from the redshift information. However due to uncertainties in
redshift estimation the redshift distribution of the members is
relatively broad. With broad cuts in redshift space one can get, on
the one
hand, a contamination of blue, non-cluster members and on the other hand,
some red cluster members might be missed due to an incorrect redshift
determination. In the previously described method
the situation is reversed. We have tested both selection criteria to
calculate the cluster luminosities (see Sect.~\ref{sc:light}) 
and both give comparable results.  

To obtain absolute rest-frame I- and R-band magnitudes for the cluster
members we determine the appropriate K-correction $K_{\rm I,R}(z)$ for
the cluster (deflector) redshift $z_{\rm d} = 0.451$ elliptical
galaxies and FORS1 filters using the GISSEL
library \citep{bruzual93} and obtain $K_{\rm I}(z_{\rm d}) =
0.378$, $K_{\rm R}(z_{\rm d}) = 0.747$. In addition we apply
galactic extinction $A_{\rm I,R}$ to the measured isophotal magnitudes
and assume zero evolutionary correction. We use $A_{\rm I}=0.121$, and
$A_{\rm R}=0.166$ from NED, where the values are obtained
from \citet{ned1} and  \citet{ned2}.

\subsection{Background galaxy catalog}
\label{sc:catweak}

In contrast to the procedure we describe above, we use aperture
magnitudes for the redshift determination of the background sources.
The reason for using aperture instead of isophotal magnitudes is that
for faint, noisy sources an estimate for the true object isophote is
hard to achieve and can bias our results for these sources.  The
diameter of the aperture is set to twice the value of the seeing given
in Table~\ref{tab:data}. In principle, one should degrade all the
images to match the seeing of the worst one (in our case U). However,
the effect is negligible compared to the photometric errors in the
U-band, and therefore we compensate for that by choosing different
sizes of the aperture.

\begin{figure*}[ht]
\begin{minipage}{17cm}
\begin{center}
\includegraphics[width = 0.7\textwidth]{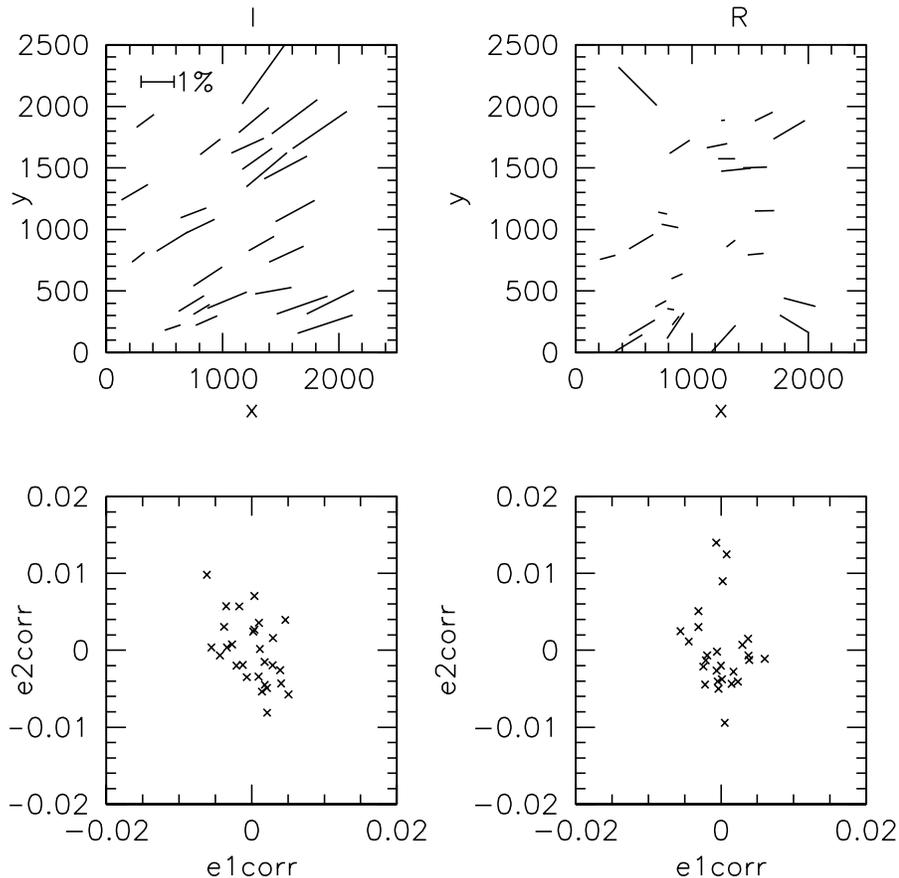}
\end{center}
\end{minipage}
\caption{The upper panels show the spatial variation of the PSF
  anisotropy in I (34 stars) and R (32 stars).  The length of the
  sticks give the amplitude of the stellar ellipticities (the length
  corresponding to $\abs{\epsilon} = 0.01$ is given at the top-left).
  The maximum ellipticity is around 2\%.  The lower panels show the
  ellipticity distribution after correction with a second-order
  polynomial. The formal residuals are about 0.005 in each component.
  Because of the small field of view and the small number of stars, the
  true errors are probably higher but difficult to estimate.}
\label{fig:PSF}
\end{figure*}

For the weak lensing analysis the R- and I-band exposures were used.
As outlined in Sect. \ref{sc:obs}, the I-band serves as our primary
weak lensing science frame. It is the image with the highest
number-density of sources that can be used for weak lensing. Below, we
cross-check our results obtained in this band with a parallel analysis
in the R-band.  We correct all galaxies in the field for the PSF
anisotropy and PSF smearing as described in \citet{erben01}. The
procedure is based on the KSB algorithm \citep{ksb95}, in particular
we use the {\tt IMCAT} implementation ({\tt
http://www.ifa.hawaii.edu/\~{}kaiser}). We select stars from the
half-light-radius vs. magnitude diagram and fit a second-order
polynomial to their measured ellipticities. In Fig.~\ref{fig:PSF} we
plot the measured PSF variation for the I- and R-band data.

\begin{figure*}[ht]
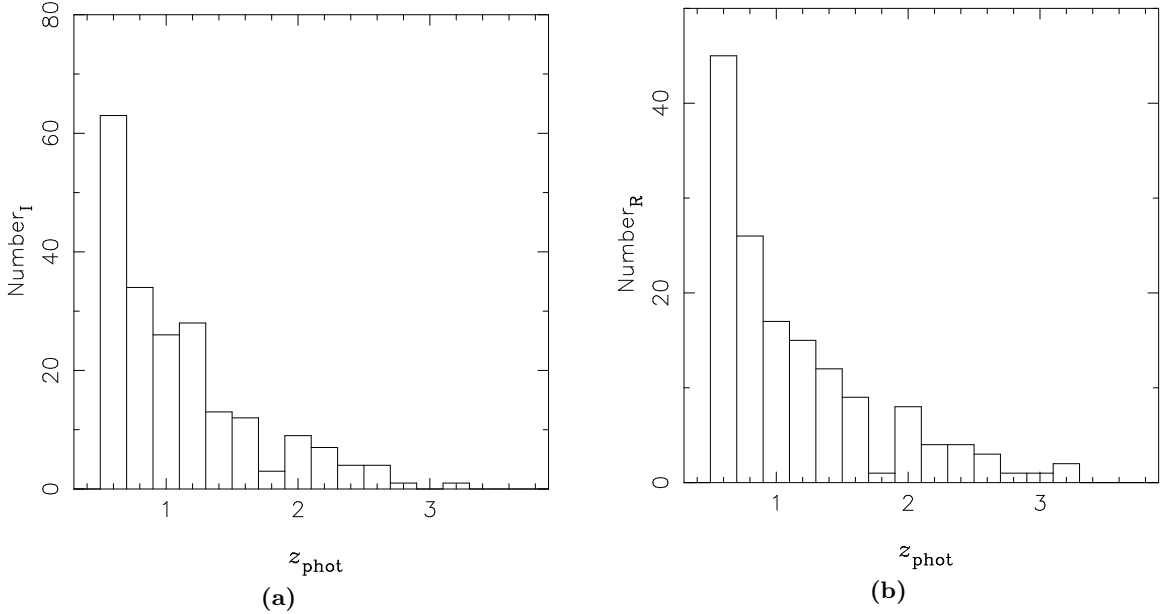

\begin{minipage}{17cm}
\begin{minipage}{8cm}
\begin{center}
\includegraphics[width = 0.9\textwidth]{2234fig3.ps}
\centerline{{\bf (a)}}
\end{center}
\end{minipage}
\begin{minipage}{8cm}
\begin{center}
\includegraphics[width = 0.9\textwidth]{2234fig4.ps}
\centerline{{\bf (b)}}
\end{center}
\end{minipage}
\end{minipage}
\caption{The redshift distributions of background sources used for
  weak lensing analysis (only sources with $z_{\rm phot}>0.55$ were
  considered) for I-band {\bf a)} and R-band catalogue {\bf
  b)}. The mean photometric redshift of the samples is $\ave{z_{\rm I}} =
  1.18$ and  $\ave{z_{\rm R}} = 1.14$.}  
\label{fig:zphot}
\end{figure*}

For the final weak lensing catalogue only sources having
photometric redshift estimate $z_{\rm phot} > 0.55$ are considered.
We end up with $N_{\rm g} = 210$ background
sources for the I-band data (giving 15 galaxies per $\mbox{arcmin}^2$), and 
with $N_{\rm g} = 140$ ($10 \mbox{ arcmin}^{-2}$) for the R-band.  The resulting
redshift distributions for both catalogues are given in
Fig.~\ref{fig:zphot}, the mean photometric redshift of the samples are
$\ave{z_{\rm I}} = 1.18$ and $\ave{z_{\rm R}} = 1.14$.

\section{Searching for multiply imaged candidates of \RXJ}
\label{sc:RXJ_strong}
\begin{figure*}[ht]
\begin{minipage}{17cm}
\begin{center}
\includegraphics[width=15cm]{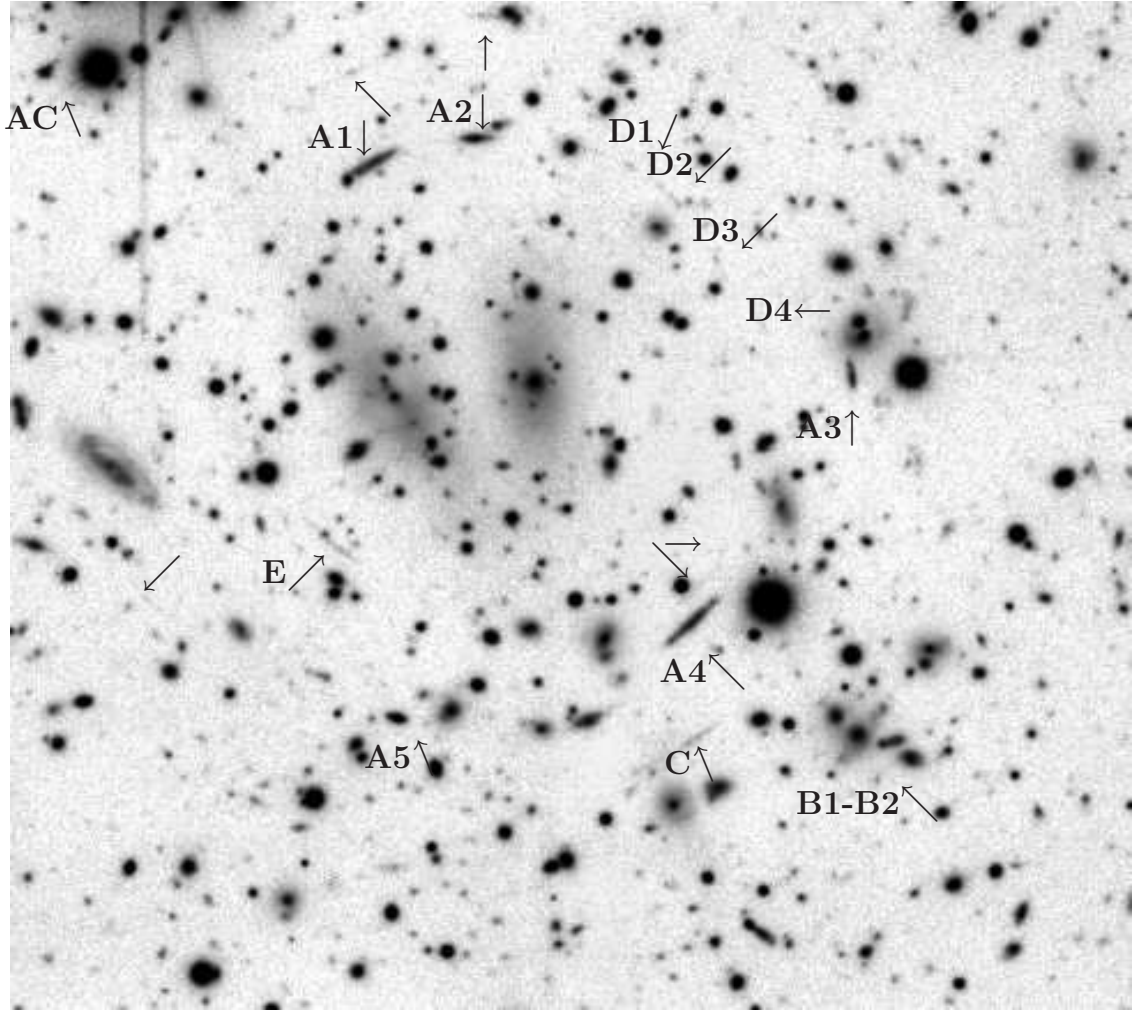}
\end{center}
\end{minipage}
\caption{The $2.7 \times 2.4  \mbox{ arcmin}^2$  
combined UBVRIKs image of the {\RXJ} showing previously
  known (system A and C) and
  newly discovered (systems B, D, and E) arc candidates (see
  Table~\ref{tab:arcs}). All are marked with a label and an arrow,
 further arc candidates are marked
  with an arrow only. The image was scaled non-linearly and the
  scaling varies across the field (in order to enhance the details of
  the image).}
\label{fig:arcs_all}
\end{figure*}

Thus far, five arc candidates for this cluster have been reported in
the literature. The first two were discovered by \citet{schindler95},
and shallow HST STIS images revealed three additional ones
\citep{sahu98}.  These five arcs (A1-A5 as labelled by
\citealp{sahu98}, see Fig.~\ref{fig:arcs_all}) are not all images of
the same source. As is obvious from Fig.~\ref{fig:RXJ_colour} most of
them have different colours and surface brightnesses. Since
gravitational lensing conserves both they belong to at least three
different sources. However, two of these arcs (A4 and A5) do have the
same colours and we consider them to be images of the same
source. Although A4 has the appearance of a very straight, edge-on
spiral galaxy (see Fig.~\ref{fig:arcs_all}), it can still be lensed,
since the cluster members to the south west of it can produce a
sufficiently strong tidal field to cause such a morphology. We note
that the arc A3 considered by \citet{allen02} to belong to this system
as well has different colours (see Table~\ref{tab:arcs_photo}).
Judging by the Fig.~\ref{fig:RXJ_colour} one would think that A1 and
A3 are multiple images of a single source as well, however the figure
is a composite of 3 bands only and the detailed photometry shows that
this is not the case(see Table.~\ref{tab:arcs_photo}).

We detect new arc candidates using the I-band and Ks-band image, as
well as the combined (following the procedure described in
\citealp{szalay99}) UBVRIKs image. In individual bands some of the arcs
could not be significantly detected. In particular, we report here on
the discovery of a red double-component arc candidate to the
south-west of A4, which we designate with labels B1 and B2. The two
components formed in the middle of a concentration of cluster
members. Their extreme red colour suggests that it is either a highly
reddened galaxy at $z\sim 1$ or it is a galaxy at $z \gtrsim 5$. In
addition we detect in the vicinity of the system B a long thin arc
candidate (C), which was also presented in \citet{lenzen04} as number
3 (see Fig.~\ref{fig:arcs}).  In the vicinity of A2 we detect
additional four arc candidates and denote them as D1-D4 (see
Fig.~\ref{fig:arcs_all}). However, we do not claim that these
components components belong to the same multiply imaged system,
although their configuration is suggestive for that.  Since these
candidates are very faint, no reliable photometry can be obtained; the
same is true for the arc candidate E. We use SExtractor to measure the
ellipticities of these arcs from the I-band (systems A, C), Ks-band
image (system B, due to its extreme red colour), and combined UBVRIKs
image (systems D, E; since they can not be significantly detected in
individual bands) -- see Table~\ref{tab:arcs}.  We detect more possible arc candidates (labelled
only with arrows in Fig.~\ref{fig:arcs_all}). They are at the limit of
the detection level and therefore their associated errors are too
large for them to be used for our analysis.

 \begin{table}[b!]
\caption{The properties of the arcs A1-A5 and the candidate
counter-image AC used in the strong lensing analysis of the cluster. We also 
present additional arc candidates (systems B, C, D, and E) -- see
also Fig.~\ref{fig:arcs_all}. 
The properties of systems A, and C are measured
from the I-band, while B is measured from the Ks-band image, and D and
E are measured from the combined UBVRIKs image. The
positions and position angles are given with respect to the brightest cluster
member; the position angle of $90^{\circ}$ means that the arc is
tangentially aligned with the BCG. All are measured with
  SExtractor. Because of the proximity of B1 and
B2 to a cluster member we can not measure their ellipticities accurately.}
\label{tab:arcs}
\begin{center}
\begin{tabular}{p{0.05\linewidth} c c c c}
\noalign{\smallskip}
\hline
\noalign{\smallskip}
\hline
\noalign{\smallskip}
Arc & $\theta_1$ [arcmin]& $\theta_2$ [arcmin] & $\abs{\epsilon}$ &
PA$[\rm deg]$\\
\noalign{\smallskip}
\hline
\noalign{\smallskip}
A1&$-0.3460$ &$\phantom{-}0.4547$& $0.571$&  $82.0$\\
A2&$-0.1253$ &$\phantom{-}0.5103$& $0.505$& $79.1$\\
A3&$\phantom{-}0.6644$& $\phantom{-}0.0090$ & $0.613$&$96.1$\\
A4 &$\phantom{-}0.3314$ & $-0.4980$ & $0.713$ & $99.8$\\
A5 &$-0.2891$ & $-0.7009$ & $0.333$ & $99.4$  \\
AC  & $-1.0241$ & $\phantom{-}0.6509$ & $0.327$ & $65.8$\\
\noalign{\smallskip}
\hline
\noalign{\smallskip}
B1 &$\phantom{-}0.7440$ & $-0.6873$ & &\\
B2 &$\phantom{-}0.6971$ & $-0.7653$ & &\\
\noalign{\smallskip}
\hline
\noalign{\smallskip}
C &$\phantom{-}0.3299$ & $-0.4985$& $0.735$ &$99.8$\\
\noalign{\smallskip}
\hline
\noalign{\smallskip}
D1  &$\phantom{-}0.1985$ & $\phantom{-} 0.4492$ & $ 0.219 $ & $81.8$\\
D2  &$\phantom{-}0.3023$ & $\phantom{-} 0.3688 $ & $0.283$ & $ 80.31$\\
D3  &$\phantom{-}0.3812$ & $\phantom{-} 0.2562 $ & $0.364 $ & $56.5$\\
D4  &$\phantom{-}0.4771$ & $\phantom{-} 0.1162$ &$0.488$&$ 106.5$\\
\noalign{\smallskip}
\hline
\noalign{\smallskip}
E &$-0.4048$& $ -0.3581$&$ 0.552$ &$108.97$\\
\noalign{\smallskip}
\hline
\end{tabular}
\end{center}
\end{table} 

 \begin{table*}[b!]
\caption{The photometric properties of the arcs A1, A2, A4, A5, and the
  candidate counter image AC. Given are
three colours ($m_{\rm B} -m_{\rm I}$, $m_{\rm V} -m_{\rm I}$, and $m_{\rm R} -m_{\rm I}$) 
in magnitudes (measured from the isophotal magnitudes), VRI peak surface
brightnesses $S_{\rm V,R,I}$ (in magnitudes), and photometric
  redshifts. For A1-A5 we determine them using 6 bands, for AC Ks is
  not available.  
If objects belong to the same source the colours and
surface brightnesses need to be conserved.}
\label{tab:arcs_photo}
\renewcommand{\footnoterule}{}  
\begin{minipage}{\textwidth}
\begin{center}
\begin{tabular}{l c c c c c c r}
\noalign{\smallskip}
\hline
\noalign{\smallskip}
\hline
\noalign{\smallskip}
& $m_{\rm B} -m_{\rm I}$  & $m_{\rm V} -m_{\rm I}$ & $m_{\rm R}
-m_{\rm I}$& $S_{\rm V}$ & $S_{\rm R}$ & $S_{\rm I}$ & $z_{\rm phot}$\\
\noalign{\smallskip}
\hline
\noalign{\smallskip}
A1 & $2.20 \pm 0.07$ & $1.78 \pm 0.05$ & $0.98 \pm 0.05$ & 24.37 & 23.75& 22.72& 0.69\\
A2 & $3.66 \pm 0.09$ & $2.51 \pm 0.05$ & $1.38 \pm 0.05$ & 24.16 & 23.44 & 21.88 & 0.73 \\
A3 & $1.52 \pm 0.07$ & $0.96 \pm 0.05$ & $0.54 \pm 0.05$ & 24.23 &
23.89 & 23.19 & 1.65\\
A4 & $0.99 \pm 0.07$ & $0.81 \pm 0.05$ & $0.53 \pm 0.05$ & 23.60 & 23.30&
22.60 & 1.76\\
A5 & $1.08 \pm 0.07$ & $0.88 \pm 0.05$ & $0.57 \pm 0.05$ & 23.60 &
23.29 & 22.60 & 1.70\\
AC & $1.28 \pm 0.07$ & $0.88 \pm 0.05$ & $0.52 \pm 0.05$ & 23.90 &
23.50 & 22.98& 1.30 \footnote{The redshift
  of AC was determined using 5 bands only. It is consistent with
  redshifts of A4 and A5 if they are also determined without Ks-band
  information.} \\
\noalign{\smallskip}
\hline
\end{tabular}
\end{center}
\end{minipage}
\end{table*}

Starting from the most plausible candidate multiple image system A4-A5
  we search for additional images belonging this system in an
  automated fashion. The aperture magnitudes of an image in either
  $N_{\rm f}=6$ or $N_{\rm f}=5$ filters $m_{i,f}$ are compared with
  the magnitudes $m_{j,f}$ of all other images in the field (where $i$
  is in our case the index of A4 or A5). We use the $\chi^2$ approach
\begin{equation}
\chi_{i,j}^2 = \sum_{f=1}^{N_{\rm f}}\frac{\rund{m_{i,f} -
    \rund{m_{{j},f} + \mu_{i,j}}}^2}{\sigma_{i,f}^2 +
    {\sigma_{j,f}^2}}\; ,
\label{eq:22}   
\end{equation}
where $\mu_{i,j}$ is the relative magnification between the images $i$
and $j$, and $\sigma_{i,f}$ and $\sigma_{j,f}$ are the magnitude
measurement errors. Since lensing is achromatic we can evaluate
$\mu_{i,j}$ by forcing $\partial \chi_{i,j}^2 / \partial \mu_{i,j} =
0$ to hold. The resulting $\chi_{i,j}^2$ function follows a
$\chi^2$-distribution with $N_{\rm f}-1$ degrees of freedom. The best
fitting images are then further visually analysed and tested for the
conservation of surface brightness. A possible counter image candidate
to A4 and A5 was found, which we label with AC. All three are
encircled dashed-yellow in Fig~\ref{fig:RXJ_colour}, their photometric
properties (and the properties of A1-A3) are listed in
Table~\ref{tab:arcs_photo}.

Using six flux measurements in UBVRIKs  for the redshift
determination of A4 and A5 and five in UBVRI for the AC (it is located at
the edge of the Ks-band image and therefore the Ks photometry is not
reliable) we find that A4 and A5 are consistent with being at a source
redshift of $z_{\rm s}\simeq1.76$. Unlike for other background objects
(see Sect.~\ref{sc:catweak}),  we use isophotal magnitudes to
obtain reliable redshifts for A4 and A5 here due to the large
ellipticity of the arcs.  The redshift determination is in agreement
with \citet{ravindranath02} who, based on the absence of the $O[II]$ line
in their spectrum, predict the redshift of A4 to be $>1.04$. The
redshift estimate of AC using 5 filters is $1.3$; however, also the
redshift estimates of arcs A4 and A5 are $1.3$ if we use only 5
filters. All three probability distributions for the redshift
estimates are very broad and the higher redshift of $1.76$ is
consistent with the photometric data in all three cases. In the
redshift regime $1.2 < z < 2$ the main features in the spectral energy
distribution (Lyman break, Balmer break, etc.) lie outside of the
optical bands and therefore the NIR photometry is important.  We
therefore use the estimated photometric redshift from UBVRIKs of
$z_{\rm s}\simeq1.76$ from now on. Unfortunately, the redshift
estimate for the multiply imaged system can substantially influence
the combined cluster mass reconstruction (the position of the critical
curve changes with redshift). We investigate this effect in
Sect.~\ref{sc:RXJ_united}.

Within the errors, the three images have the same
colours as well as the same surface brightnesses (see
Table~\ref{tab:arcs_photo}). In addition, the
photometric redshift estimate (using 6 filters) is the same for A4
and A5. The colours and peak surface brightnesses of the counter image
are also consistent, however due to its smaller apparent size its
photometry is less reliable. There are more candidate multiple image
systems in this field; they will be the subject of a future study.

\section{Cluster mass reconstruction of \RXJ}
\label{sc:RXJ_rec}
In this section we present the mass modelling of the cluster {\RXJ}.
We first give a short outline of the method, a full account of it can
be found in \citetalias{bradac04a}.

\subsection{Short outline of the method}
\label{sc:outline}
The main idea behind the method is to parametrise the
cluster mass-distribution by a set of model parameters, where this
parametrisation is chosen as generic as possible. In our case we use
the gravitational potential $\psi$ on a regular
grid.  We factorise
the redshift dependence by the so-called ``cosmological weight''
function $Z(z)$, as defined in \citetalias{bradac04a} (see also \citealp{bartelmann00}).\footnote{To evaluate the angular diameter distances we assume
 the $\Lambda$CDM cosmology with $\Omega_{\rm m} = 0.3$, $\Omega_{\Lambda} =
0.7$, and Hubble
constant $H_0 = 70 {\rm km \:
  s^{-1}\:Mpc^{-1}}$.}

We define the $\chi^2$-function 
\begin{equation}
  \label{eq:13a}
  \chi^2(\psi_k) = \chi_{\epsilon}^2(\psi_k)  + \eta R(\psi_k) + \chi_{\rm M}^2(\psi_k) \; .
\end{equation}
where $\chi_{\epsilon}^2(\psi_k)$ is the contribution from
weak lensing and $\chi_{\rm M}^2(\psi_k)$ from strong lensing. In
addition, the regularisation $R(\psi_k)$ with regularisation
parameter $\eta$ is employed in order to penalise any models that
would follow the noise pattern in the data. We minimise the $\chi^2$
function with respect to $\psi_k$ by solving the equation $\partial
\chi^2(\psi_k) / \partial \psi_k = 0$. This is in general a non-linear
set of equations, and we solve it in an iterative manner. We linearise
this system and starting from some trial solution
(i.e. $\kappa^{(0)}$, $\gamma^{(0)}$, and $\vec \alpha^{(0)}$) we
repeat the procedure until convergence is achieved.   We showed in
\citetalias{bradac04a} that different models used as a trial solution
do not influence results significantly. We try to confirm this
result by investigating different models here as well.

\subsection{Initial conditions for the method}
\label{sc:RXJ_ini}

\begin{figure*}[ht]
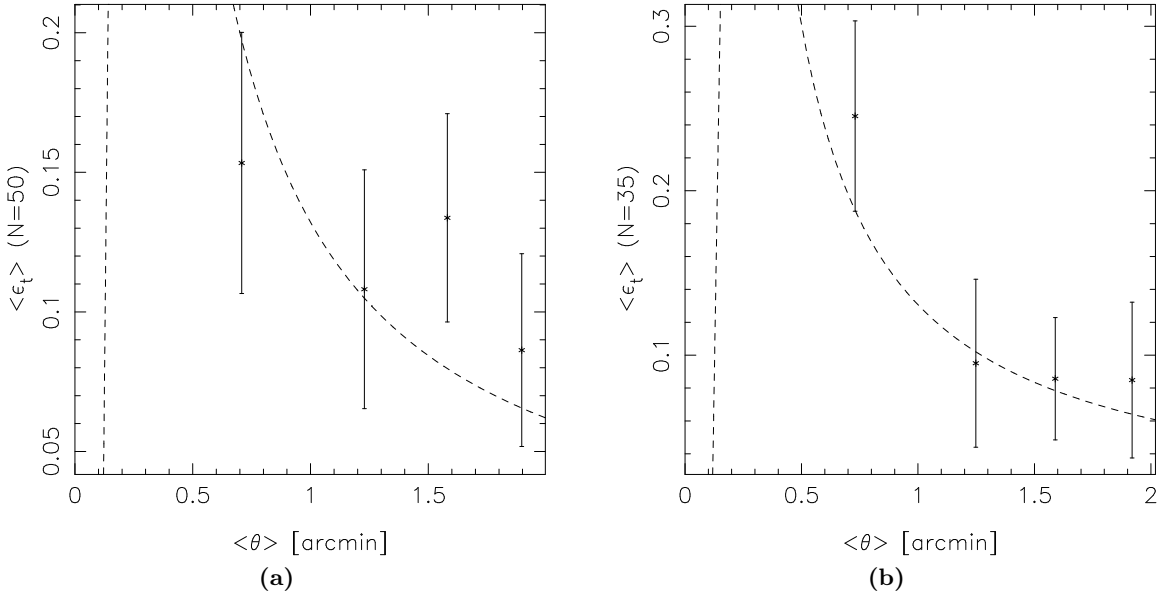

\begin{minipage}{17cm}
\begin{minipage}{8cm}
\begin{center}
\includegraphics[width = 0.9\textwidth]{2234fig5.ps}
\centerline{{\bf (a)}}
\end{center}
\end{minipage}
\begin{minipage}{8cm}
\begin{center}
\includegraphics[width = 0.9\textwidth]{2234fig6.ps}
\centerline{{\bf (b)}}
\end{center}
\end{minipage}
\end{minipage}
\caption{Average tangential ellipticity $\ave{\epsilon_{\rm t}}$ vs. projected
  radius $\ave{\theta}$ in radial bins centred on the brightest
  cluster member containing 50 galaxies per bin for the I-band data {\bf
    a)} and 35 galaxies for the R-band data {\bf b)}. The errors are
  obtained by randomising the phases of the measured ellipticities,
  while preserving their absolute values. 100 randomisations were
  performed. The dashed line shows the best-fit SIS profile to the
  unbinned data, for the I-band we obtain $\sigma_{\rm I, SIS} = 950
  \pm 60 \mbox{ kms}^{-1}$ and for the R-band $\sigma_{\rm R, SIS} =
  960 \pm 70 \mbox{ kms}^{-1}$ (both $1\sigma$ error bars). It is
  plotted here for the average source redshifts of $\ave{z_{\rm I}} = 1.18$ and
  $\ave{z_{\rm R}} = 1.14$. In the interval  $0^{\prime} < \ave{\theta}
  \lesssim 0.\arcminf25$ the expectation value of the observed ellipticity
  (for these models) is given by $1/g(\theta,z)^{*}$ and 
the tangential ellipticity profile has 
a steep gradient at $\ave{\theta} \simeq 0.\arcminf1$
  (for both bands) where $Z(\ave{z_{\rm R, I}}) \kappa = 1$ [since
  $g=Z\gamma / \rund{1 - Z\kappa}$]. At
  $\ave{\theta}\gtrsim  0.\arcminf25$ the expectation value of 
 the observed ellipticity is given by $g(\theta,z)$.}
\label{fig:RXJ_wlprof}
\end{figure*}

For the purpose of obtaining the initial values for $\kappa^{(0)}$, 
$\gamma^{(0)}$, and  $\vec \alpha^{(0)}$ we first investigate the signal 
from the averaged tangential ellipticities and fit these using the singular
isothermal sphere SIS model (hereafter called {\it IS} scenario). The
tangential ellipticities are given by
\begin{equation}
\epsilon_{\rm t} = -\Re\eck{\epsilon \:\rm
  e^{-2{\rm i}\phi}}
\label{eq:21}
\end{equation}
where $\phi$ specifies the direction to the source galaxy with respect
to the
the brightest
cluster galaxy (BCG). In
Fig.~\ref{fig:RXJ_wlprof} we plot the average tangential ellipticities
versus projected radius $\ave{\theta}$ in radial bins centred on the BCG, 
containing 50 (35)  galaxies each. Both, the I- and R-band data
give comparable results. We note that the tangential ellipticity
signal is still high at the edge of the field $\ave{\epsilon_{\rm t}} \sim
0.1$, thus making this data inaccessible for standard weak lensing
techniques aiming to determine the mass, since 
on this relatively small field one
cannot break the mass-sheet degeneracy by simply assuming
$\kappa \sim 0$ at the field edges.

We fit an SIS profile to the individual tangential
ellipticities (not binned), the model ellipticities are 
calculated using the redshifts of these sources.
The resulting line-of-sight velocity dispersion is $\sigma_{\rm I,
  SIS} = 950 \pm 60 \mbox{ kms}^{-1}$ for the I-band data and
$\sigma_{\rm R, SIS} = 960 \pm 70 \mbox{ kms}^{-1}$ for the R-band
(both $1\sigma$ error bars). The tangential ellipticity as a function
of radius for this model is plotted in Fig.\ref{fig:RXJ_wlprof}
(dashed line) for the average source redshifts of $\ave{z_{\rm I}} =
1.18$ and $\ave{z_{\rm R}} = 1.14$ (see Sect.~\ref{sc:catweak}).  In
addition, the absence of the lens is excluded with more than
$10\sigma$ significance in both bands (all minimisations and error
analysis in this subsection are performed using \texttt{C-minuit} from
\citealp{james75}).

The line-of-sight velocity dispersion estimates are higher than the
measured velocity dispersion from \citet{cohen02}, and lower than
previous weak lensing, strong lensing and X-ray measurements. However,
in the optical it is evident that the cluster has a lot of structure
and therefore the SIS profile does not describe the cluster
adequately. It has at least two main components; in addition there is
X-ray emission off-centred from the BCG. Furthermore, at the scales
where we measure the profile, $\lesssim 400 {h^{-1}}\mbox{ kpc}$, the
profile of the cluster is probably not isothermal (see e.g.
\citealp{navarro04}). Therefore, the values of $\sigma$ obtained in
this manner should not be trusted, we only use them for one of the
initial models for $\kappa^{(0)}$

Another possibility to obtain initial conditions is to use the
multiple image information for the cluster. We perform a very rough
analysis by using the data for the arc system A4-A5-AC (given
in Table~\ref{tab:arcs}). In addition to the
image positions we also use image ellipticities as constraints. The
model consists of 
a non-singular isothermal ellipse (NIE) \citep{ke98}, given by 
\begin{equation}
  \kappa(\vec {\theta'}) = \frac{b_0}{2\sqrt{\frac{1+\abs{\epsilon_{\rm
            g}}}{1-\abs{\epsilon_{\rm g}}} \rund{r_{\rm c,nis}^2 +
        (\theta'_1)^2} + (\theta'_2)^2}} \; ,
\label{eq:nis}
\end{equation}
where $\vec \theta'$ is calculated w.r.t. the major axis of the cluster
surface mass density.  We allow the centre of the cluster potential
$\vec \theta_{\rm cl}$, the scaling $b_0$, ellipticity
$\abs{\epsilon_{\rm g}}$, and the position angle $\phi_{\rm g}$ to
vary.  Following the prescription of \citet{kneib96} we also include
the 10 brightest cluster members in the I-band to the mass model.
They are modelled as non-singular isothermal spheres with a
line-of-sight velocity dispersion $\sigma_{\rm nis}$ and core radius
$r_{\rm c,nis}$ following
\begin{align}
\sigma_{\rm  nis} & {}  \propto L^{1/4} \; ,& r_{\rm c,nis}  & {} \propto
L^{1/2} \; ,& 
   \label{eq:23}
 \end{align}
 The proportionality constants were chosen such that the I-band
 magnitude $m_{I}=17.5$ galaxy would have $\sigma_{\rm sis} = 300
 \mbox{ km\,s}^{-1}$ and $r_{\rm c,nis} = 0.\arcminf1$ (the BCG has
 $m_{I}=17.8$). We also fix the core radius of the cluster to
 $0.\arcminf3$. These constants are {\it not} allowed to vary.  The
 best fit model for this system has values of $\{\theta_{{\rm cl},1},
 \theta_{{\rm cl},2}, b_0, \abs{\epsilon_{\rm g}}, \phi_g\} =
 \{-0.\arcminf21,-0.\arcminf10,0.\arcminf97,0.3,0.8\}$.

We stress here that it was not our aim to obtain a detailed strong
lensing cluster-mass model, since it will only be used for the initial
values of reconstruction. The multiple image system used here is
independently included in the non-parametric reconstruction.  We have
shown in \citetalias{bradac04a} (and also confirm this in
Sect.~\ref{sc:RXJ_united}) that the reconstruction depends little upon
the details of the initial model we use; for this reason a detailed
modelling is not needed. In particular, the precise choice of those
parameters that we did not vary in the modelling is not very relevant
in our case.  For the same reason we also do not include additional
multiply imaged candidates in the analysis.

\begin{figure*}[ht]
\begin{center}
\begin{pspicture}(0,0)(\textwidth,1.08\textwidth)
\rput[lb](0,0){\includegraphics[width=1.0\textwidth]{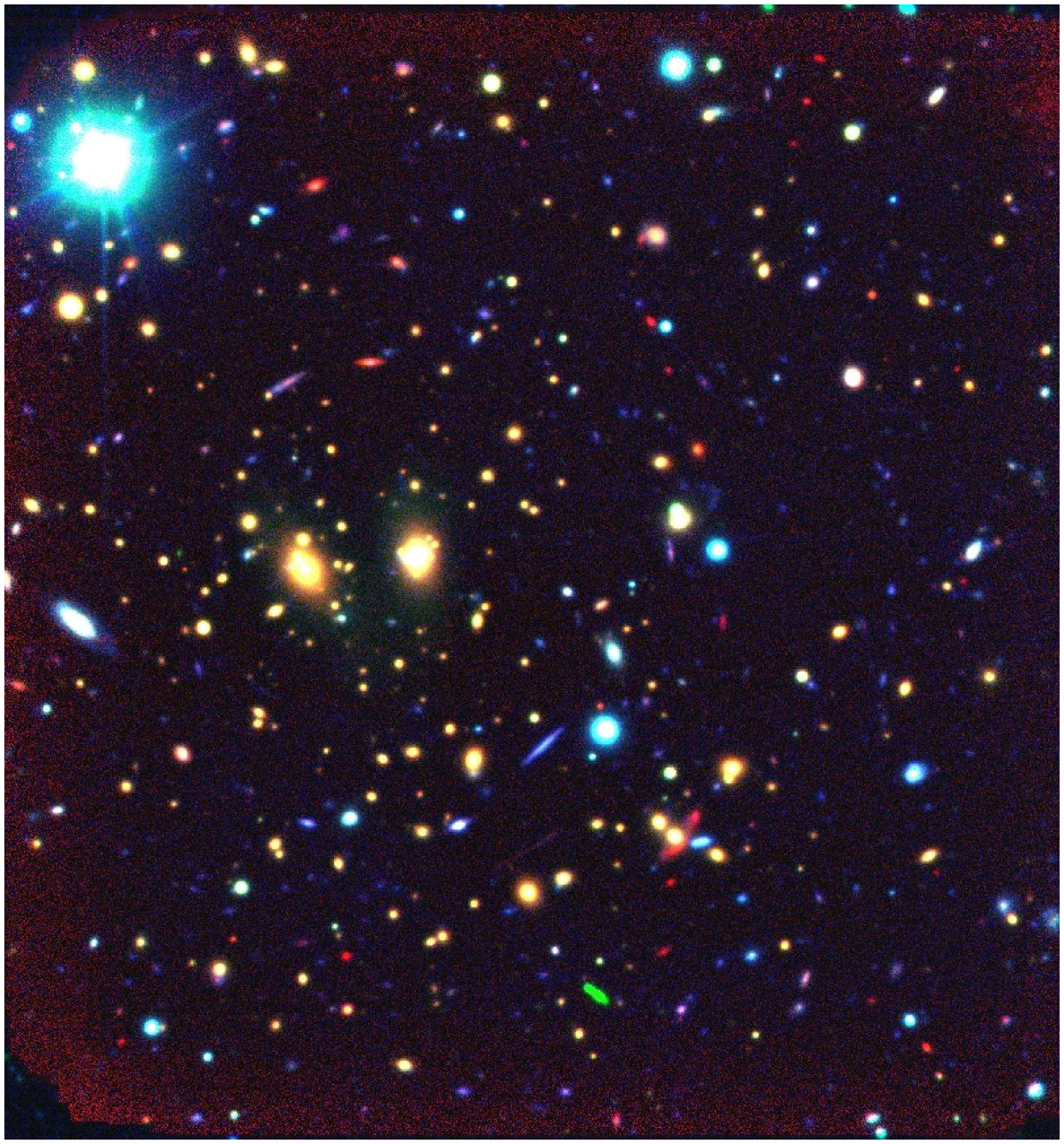}}%
\rput[lb](0,0){\includegraphics[width=1.0\textwidth]{2234fig8.ps}}%
\rput[lb](0,0){\includegraphics[width=1.0\textwidth]{2234fig9.ps}}
\end{pspicture}
\end{center}
\caption{The BRK colour composite of the  $\sim 3 \times 3.2 \mbox{
  amin}^2$ field of {\RXJ}. Overlaid is in white contours the combined weak and strong
  lensing mass reconstruction from Fig.~\ref{fig:RXJ_wl}a1.  The
  contour levels are the same (the field here is smaller), we 
smooth them here using a Gaussian kernel
characterised by $\sigma=5^{\prime\prime}$ for clarity of the plot. 
Yellow circles show the multiple image system we use. North is
  up and East is left.}
\label{fig:RXJ_colour}
\end{figure*} 

\begin{figure}[ht]
\begin{minipage}{8.5cm}
\begin{center}
\includegraphics[width=8cm]{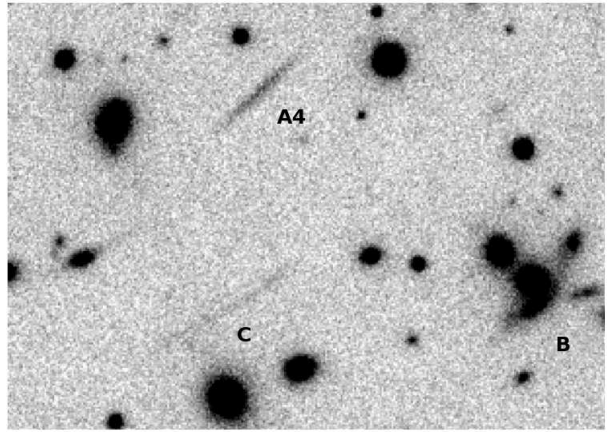}
\end{center}
\end{minipage}
\caption{A $\sim 1.\arcminf0 \times 0.\arcminf6$ 
cutout of the Ks-band image of {\RXJ} showing one of
  the arcs used for strong lensing, the newly discovered 
very red arc system B1-B2 (two images almost merging,
c.f. Fig.~\ref{fig:RXJ_colour}) and  
the long thin arc C.}
\label{fig:arcs}
\end{figure}

\subsection{Combined weak and strong lensing mass reconstruction of {\RXJ}}
\label{sc:RXJ_united}

\begin{figure*}[ht]
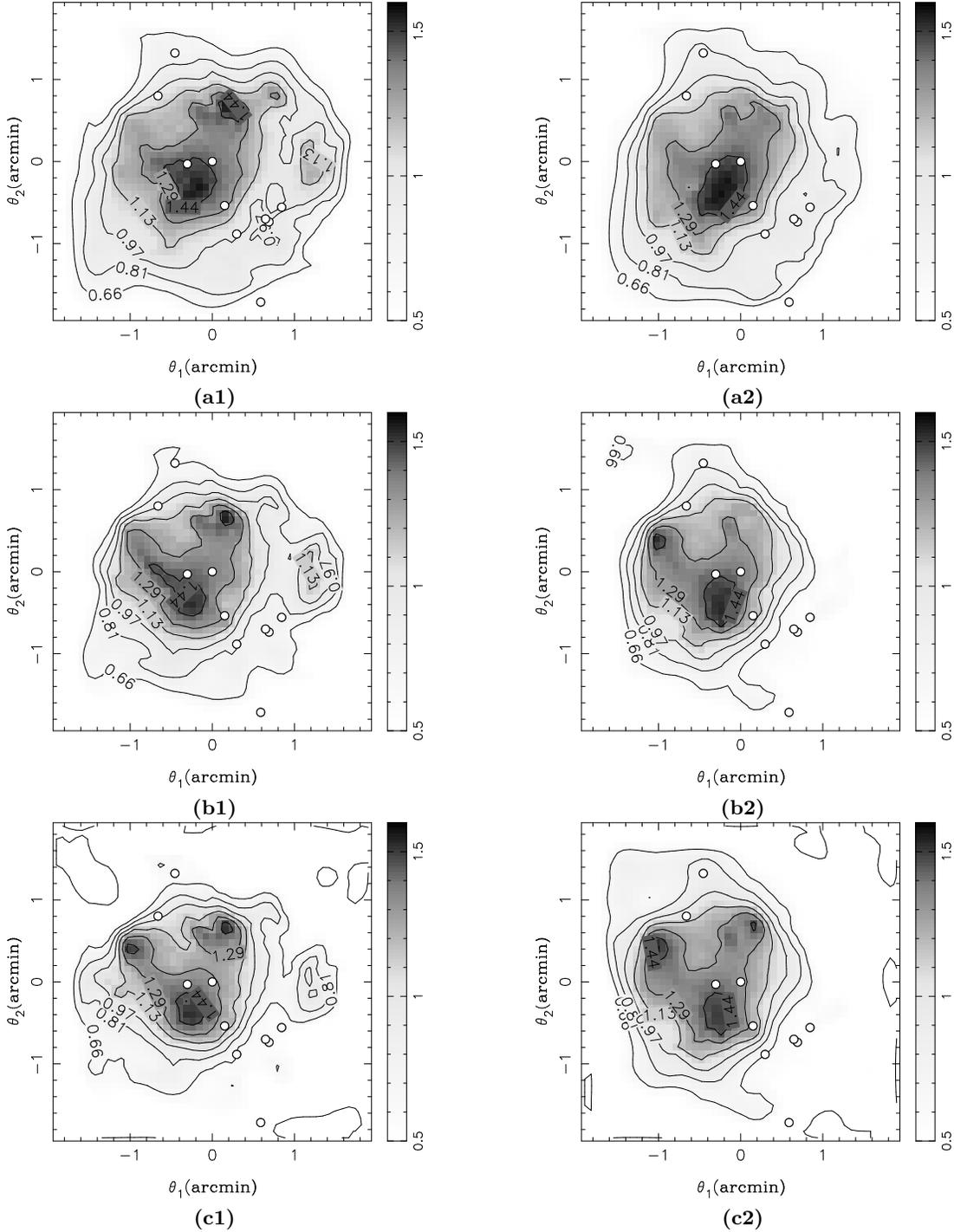

\begin{minipage}{17cm}
\begin{minipage}{8cm}
\begin{center}
\includegraphics[width=0.8\textwidth]{2234fig12.ps}
\centerline{{\bf (a1)}}
\includegraphics[width=0.8\textwidth]{2234fig13.ps}
\centerline{{\bf (b1)}}
\includegraphics[width=0.8\textwidth]{2234fig14.ps}
\centerline{{\bf (c1)}}
\end{center}
\end{minipage}
\begin{minipage}{8cm}
\begin{center}
\includegraphics[width=0.8\textwidth]{2234fig15.ps}
\centerline{{\bf (a2)}}
\includegraphics[width=0.8\textwidth]{2234fig16.ps}
\centerline{{\bf (b2)}}
\includegraphics[width=0.8\textwidth]{2234fig17.ps}
\centerline{{\bf (c2)}}
\end{center}
\end{minipage}
\end{minipage}
\caption{$\kappa$-maps obtained from combined strong and weak lensing
reconstruction of the cluster \RXJ. {\bf Left} panels show the
reconstructions using I-band data and for the ones on the {\bf right}
we use the R-band data. For the I-band data we have $N_{\rm g} = 210$
background source galaxies, and for the R-band $N_{\rm g} = 148$, all
with known photometric redshifts. In {\bf a1)} and {\bf a2)} we use
the best fit model from the strong lensing analysis of the cluster
{\it IM} as initial condition, in
{\bf b1)} and {\bf b2)} we use the {\it IS} model (SIS model fitted to
tangential ellipticities, centred
on the brightest cluster member) -- for both {\it IM} and  {\it IS}
see Sect.~\ref{sc:RXJ_ini}. In {\bf c1} and {\bf c2)} we use
{\it I0}, i.e.  $\kappa^{(0)}=0$ on all grid points.  The positions of
the 10 brightest cluster members are plotted as white circles.}
\label{fig:RXJ_wl}
\end{figure*}

We apply the mass-reconstruction method to the I- and R-band data of
{\RXJ} and the strong lensing system A4-A5-AC only (see
Table~\ref{tab:arcs}). We use three different initial models for
$\kappa^{(0)}$; {\it IM} is the best fit model from the strong lensing
analysis of the cluster, the {\it IS} model is the best fit SIS model
to binned tangential ellipticities (centred on the brightest cluster
member) -- both presented in Sect.~\ref{sc:RXJ_ini} -- and {\it I0}
has $\kappa^{(0)} = 0$ ($\gamma^{(0)} = 0$, $\vec\alpha^{(0)} = \vec
0$).  The initial regularisation parameter is set to $\eta = 400$ for
the I-band and $\eta = 200$ for the R-band. It is adaptively adjusted
in each iteration step such that the resulting $\chi^2 / N_{\rm g}
\sim 1$.  The resulting $\kappa$-maps are given in
Fig.~\ref{fig:RXJ_wl}. We also overlay the contours from
Fig.~\ref{fig:RXJ_wl}a1 to the colour composite image in
Fig.~\ref{fig:RXJ_colour}.

We estimate the mass within the cylinder of a radius of $1.\arcminf5$
(for the cluster redshift $z_{\rm d} = 0.451$ this corresponds to $360
{h^{-1}}\mbox{ kpc}$), the estimates are given in
Table~\ref{tab:mass_RXJ}. The projected mass of the cluster is
estimated to be $M(<360\:
{h}^{-1}\mbox{kpc})= (1.2 \pm 0.3) \times 10^{15} M_{\odot}$. The
error was estimated by bootstrap resampling the background galaxies in
the weak lensing catalogues. This means that for each catalogue we
randomly select $N_{\rm g}$ galaxies with replacement, if a galaxy is
selected twice (or more) we assign double (or multiple) weight to its
$\chi_{\epsilon}^2$ contribution. We generate 10 new catalogues and
perform a new mass reconstruction; the error is then given as the
variance of these estimates.  It is larger than what we obtain from
simulations in \citetalias{bradac04a}, which is partly attributed to
the fact that we only use a three-image and not a four-image system
here.  However, within the given errors the results for both bands and
for different initial models are consistent.

The projected mass from XMM measurements \citep{gitti04} within a
cylinder of the same radius as we use is given by $M_{\rm X}(<360\:
{h}^{-1}\mbox{kpc})= (0.7 \pm 0.2) \times 10^{15} M_{\odot}$ (Myriam
Gitti, private communication).  The resulting mass from the strong and
weak lensing mass reconstruction is higher and marginally consistent
with X-ray measurements. If the mass estimate is extrapolated at
larger radii (assuming an isothermal profile) it is also consistent
with the previous weak-lensing results by \citet{fischer97}. It is
however significantly larger than the mass estimate obtained by the
velocity dispersion measurement of \citet{cohen02}.

A possible explanation for the discrepant dynamical mass estimates was
presented by \citet{cohen02}. They argue that the cluster is most
likely in a pre-merging process (with clumps merging preferentially in
the plane of the sky). In such a scenario, until the merging is
complete and the cluster is virialised, the dynamical cluster mass
will be largely underestimated. On the other hand the X-ray
temperature can be increased in such merging processes (thus the mass
would be overestimated) and for this reason the south-east quadrant is
excluded (the surface brightness profile is determined by averaging
data only in the other three quadrants) in the X-ray analysis. The
temperature measurements from \citet{gitti04} thus further supports
the merger hypothesis. However if there is some extra mass present in
the excluded quadrant (as suggested by our mass maps), the mass
estimate obtained in this way from X-rays will be underestimated. If
the hypothesis is correct, traditional mass estimates relying on
equilibrium assumptions fail and gravitational lensing (with high
quality data) provides the most accurate estimate for the cluster
mass.

We note, however, that our results depend upon the correct
redshift determination and identification of the members of the
multiple image system we use. If we put the multiple image system to a
redshift of $\sim 3$ ($\sim 1.3$), the estimated mass
decreases (increases) by $\sim 10\%$. If the images do not belong to
the same system, the changes might be even more drastic. However, at
least for the two arcs A4 and A5, based on their photometric
properties, we consider this possibility less
likely (see Sect.~\ref{sc:RXJ_strong}). As a test we have also performed the 
reconstruction using only the two arcs A4 and A5. The results remain
unchanged, however the scatter between the three initial models and
the errors are larger by a factor $\sim 2$.

Further, the results rely on the correct determination of
the photometric redshifts for the weak lensing sources. The random
error of the determination is not crucial, the problem are the
systematic uncertainties. It is not excluded that e.g. some foreground sources
get assigned a high redshift and thus diluting (if they are
randomly oriented) or enhancing the signal (if they are aligned). In
addition, outliers can have $Z(z)$ assigned which is very different to
their real cosmological weight. These outliers were considered in
\citetalias{bradac04a}, they were chosen at random and their
  fraction was taken to be 10\%. Still, their
presence did not significantly change our conclusions. If, however,
their fraction is higher and/or more importantly if 
they bias the final redshift distribution, this can bias our mass estimate.

An additional test for the accuracy and reliability of our model could
  be performed by using its predictive power. Namely, if the model is
  well constrained it should be capable of predicting the position of
  e.g. the counter image to the arc A1 (providing its redshift
  determination is correct). We have tested our models using the
  following procedure.  Using the resulting potential from strong and
  weak lensing reconstruction we project (using bilinear interpolation
  and finite differencing) the position $\vec \theta_{\rm s}$ of an
  arc candidate (e.g. A1) back to the source plane and denote the
  resulting position as $\vec y_{\rm s}(\vec \theta_{\rm s})$. Then we
  search for all possible solutions $\vec \theta$ satisfying the
  non-linear set of equations $\vec y(\vec \theta) - \vec y_{\rm
  s}(\vec \theta_{\rm s}) = \vec 0$. These should then lie close to
  the possible counter image candidate(s). However, since our model is
  tightly constrained only in the vicinity of multiple images we use
  (A4, A5, and AC), the scatter of possible solutions is large. This
  issue could however be easily resolved in the future with e.g. ACS
  observations, since many more arc candidates will be found and their
  morphology can be obtained allowing for unambigous identification of
  multiple imaged systems and tighter constraints of the model.

 \begin{table}[b!]
\caption{Reconstructed mass of 
{\RXJ} within a cylinder of $360 {h^{-1}}\mbox{ kpc}$ radius around the BCG 
from I-band (left) and R-band (right) weak lensing
  data and one candidate 3-image system. Three different
  $\kappa^{(0)}$ models have been used. We use the best fit
  model from the multiple image system  {\it IM}, {\it
    IS} is the best fit SIS model from the process of parametrised
  fitting of weak lensing data and {\it
    I0} has  
  $\kappa^{(0)}=0$ on all grid points (see Sect.~\ref{sc:RXJ_ini}). 
In brackets we give for
  comparison the velocity
  dispersion of an SIS having the same enclosed mass  within
   $360 {h^{-1}}\mbox{ kpc}$.}
\label{tab:mass_RXJ}
\begin{center}
\begin{tabular}{p{0.05\linewidth} c c  c c}
\noalign{\smallskip}
\hline
\noalign{\smallskip}
\hline
\noalign{\smallskip}
 & $M_{I}$&$ [\sigma_{\rm I,SIS}]$ &
 $M_{R}$&$ [\sigma_{\rm R,SIS}]$\\
 & $[10^{15} M_{\odot}]$&$ [{\rm km s^{-1}}] $& $[10^{15} M_{\odot}] $&$ [{\rm km s^{-1}}]$\\
\noalign{\smallskip}
\hline
\noalign{\smallskip}
{\it IM} & $1.37 \pm 0.04 $& [1900]&$1.31 \pm 0.03$& [1860]\\
 {\it IS} & $1.2 \pm 0.1$ & [1800] & $1.1 \pm 0.1$& [1700]\\
 {\it I0} & $1.1 \pm 0.1$ & [1700] & $1.1 \pm 0.1$& [1700]\\
\noalign{\smallskip}
\hline
\end{tabular}
\end{center}
\end{table} 

\subsection{Rest-frame I- and R-band brightness distribution and
  mass-to-light ratio of
  {\RXJ}}\label{sc:light}
\begin{figure*}[ht]
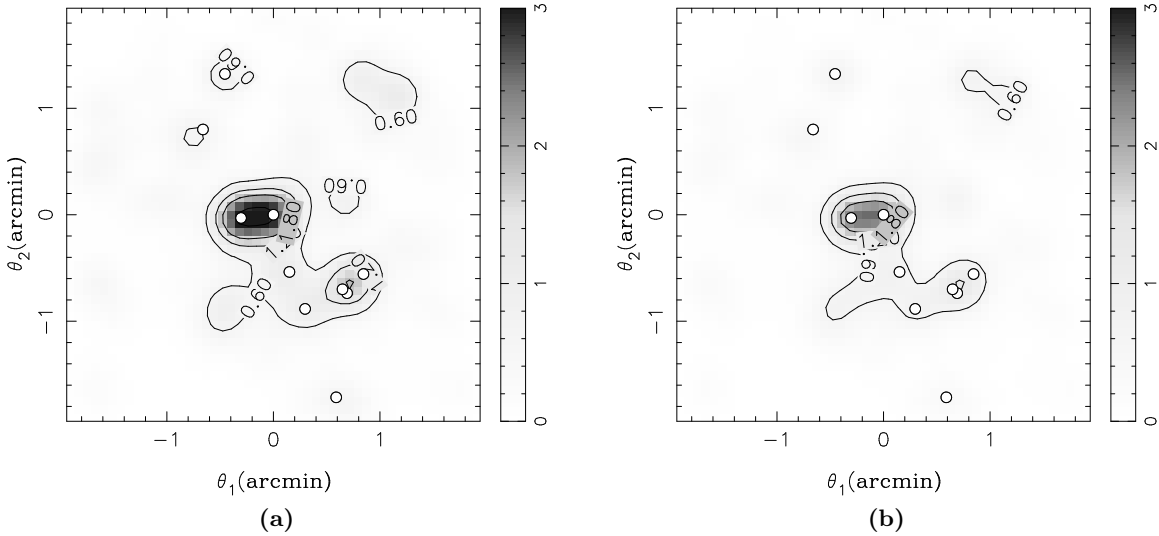

\begin{minipage}{17cm}
\begin{minipage}{8cm}
\begin{center}
\includegraphics[width = 0.9\textwidth]{2234fig18.ps}
\centerline{{\bf (a)}}
\end{center}
\end{minipage}
\begin{minipage}{8cm}
\begin{center}
\includegraphics[width = 0.9\textwidth]{2234fig19.ps}
\centerline{{\bf (b)}}
\end{center}
\end{minipage}
\end{minipage}
\caption{The I-band {\bf a)} and R-band {\bf b)} 
brightness distribution of the {\RXJ} in
  $10^{12}L_{\odot} / \mbox{ arcmin}^2$. The cluster members are
selected using the colour cuts described in
the text, the ten brightest ones (in I) are plotted as white circles. 
Their luminosities of the cluster members have 
been smoothed using a Gaussian kernel 
characterised by $\sigma=9^{\prime\prime}$.}
\label{fig:light}
\end{figure*}

To obtain the cluster brightness distribution and aperture luminosity
we proceed as follows. Using colour and redshift information for 
the selection criteria described in
Sect.~\ref{sc:catclusters} we determine the 
luminosities of the cluster members in the field. They  
are smoothed using a Gaussian kernel characterised by
$\sigma=9^{\prime\prime}$, resulting in the brightness distribution
shown (for I-band only) in Fig.~\ref{fig:light}. We then 
determine the aperture luminosity $L_{\rm a}$ by adding
the luminosities of the cluster members within $360 {h^{-1}}\mbox{
kpc}$ radius around the BCG.

The resulting I- and R-band aperture luminosities
are $L_{\rm a,I}(<360 {h^{-1}}\mbox{kpc}) = 3.1 \times 10^{12}
L_{\odot}$ and  $L_{\rm a,R}(<360 {h^{-1}}\mbox{kpc}) = 2.2 \times
10^{12} L_{\odot}$, respectively. The mass-to-light ratios (M/L) are
$M/L_{\rm I} = 400 \pm 150  M_{\odot} /  L_{\rm I,\odot}$ and 
$M/L_{\rm R} = 550 \pm 150  M_{\odot} /  L_{\rm R,\odot}$.
The cluster has only 300 members across the observed field, 
it is under-luminous in optical bands. 
In addition, 
we are measuring the M/L ratio in the inner part of the cluster,
which might not reflect the M/L ratios measured out to
$\sim1\mbox{ Mpc}$ distances from cluster centres usually quoted in
the literature.

The first concern with luminosity estimates is completeness. For this
purpose we fit the Schechter luminosity function \citep{schechter76}
to the cluster member counts as a function of absolute magnitudes
$M_{\rm I}$ and $M_{\rm R}$.  The resulting best-fit characteristic
magnitudes are $M^{*}_{\rm I} = -22.4 \pm 0.2$ and $M^{*}_{\rm R} =
-21.7 \pm 0.3$ and the faint end slopes are given by $\alpha_{\rm
  I}=-0.89 \pm 0.06$, and $\alpha_{\rm R}=-0.9 \pm 0.1$. We conclude
that our catalogues are complete to a magnitude $M^{*}_{\rm I,R}-4$
and therefore the contribution from incompleteness is negligible.

A more severe concern is the contamination by non-cluster members and
rejection of the actual members. In order to check against this, one
needs to investigate the galaxy population ``outside'' the cluster
region (on images taken with the same photometric conditions and depth
as the images we use). A slightly different approach for the purpose
of the M/L calculations can be followed by defining an outer aperture
at the edge of the image and subtracting the luminosity density in
that aperture from that in the inner portions of the cluster. The same
approach needs to be undertaken when calculating the mass. If the M/L
is constant across the field, this would give its correct value.
Unfortunately our observed fields span only $\sim 450\mbox{
  kpc}\:h^{-1}$ around the brightest cluster galaxy and therefore this
approach is not reliable. We conclude that the error budget on
luminosity is dominated by the systematics of the cluster member
selection and contamination and is very difficult to estimate. We
investigate two different selection criteria for cluster members in
Sect.~\ref{sc:catclusters}; we used colour information as well as the
photometric redshifts. The aperture luminosities from these two
criteria are consistent at the $5\%$-level. These two approaches share
similar systematics, both use the same magnitude measurements, and for
the colour-colour selection blue galaxies are added using photometric
redshift measurements.  However, in order to estimate the M/L the mass
determination is a dominant source of error.

\section{Conclusions}
\label{sc:RXJ_concl}

The case of {\RXJ} has been a cause of many puzzles in the past. Very
discrepant mass estimates are given in the literature, and
unfortunately this cluster is not the only case where the mass
measurements have proven to be difficult. We have applied a new mass
reconstruction method to deep optical data using a multiple-image
system with three images selected based on their colours and
redshifts. Our main conclusions are the following.
\begin{enumerate}
\item The combined strong and weak lensing mass
reconstruction confirms that the most X-ray luminous cluster is indeed
very massive. If the redshift and identification of the multiple-image
system, as well as redshifts used in weak lensing data, are correct we 
estimate the enclosed cluster mass within $360\: {
  h}^{-1}\mbox{kpc}$ to $M(<360\: {h}^{-1}\mbox{kpc})= (1.2 \pm
  0.3) \times 10^{15} M_{\odot}$.
\item  The reconstruction shows a south-east mass extension, 
compatible with the X-ray measurements (see e.g. \citealp{gitti04}, \citealp{allen02}). 
\item A single SIS fit to the average tangential ellipticities 
does not give a reliable estimate for the enclosed mass within $360\: {
  h}^{-1}\mbox{kpc}$;
detailed modelling needs to be performed.
\item We have demonstrated the feasibility of breaking the mass-sheet
  degeneracy in practice by using shape measurements and adding the
  information on individual redshifts, without any assumptions
  regarding the cluster potential.
\end{enumerate}
In addition we measured the corresponding mass-to-light ratio of the
cluster within $360\:{h}^{-1}\mbox{kpc}$. We find that the cluster is
more luminous in the rest-frame I-band than R-band, which is expected due to the
presence of many old (red) elliptical galaxies in clusters. The
resulting mass-to-light ratios are high, both in rest-frame I- and
R-band, giving $M/L_{\rm I} = 400 \pm 150 M_{\odot} / L_{\rm I,\odot}$
and $M/L_{\rm R} = 550 \pm 150 M_{\odot} / L_{\rm R,\odot}$. These
values are higher than typical values for clusters claimed in the
literature ($\sim 200$ for R-band). However it is difficult to compare
our results with existing measurements of the mass-to-light ratios,
since they are usually performed at larger radii not accessible with
our data.

In the course of this research we discovered one new extremely red arc 
candidate (system B) at $\sim 1\mbox{ arcmin}$ distance from the BCG.
Unfortunately its redshift can not be measured, as it is significantly
detected only in the Ks band. Further arc candidates are discovered
from the combined colour image, suggesting that the cluster is indeed
very centrally concentrated.  In addition, the enclosed mass we obtain
using the combined reconstruction also fits reasonably well the
standard mass vs. X-ray luminosity relation (see
\citealp{reiprich02}), provided we assume the model to be
  isothermal (which for the same enclosed mass as our reconstructed
  model means
$\sigma \simeq 1800\mbox{ kms}^{-1}$) to a radius of $r_{200}$
frequently used to determine the relation.

The mass-reconstruction of {\RXJ} can be significantly improved. Deep
HST imaging would greatly help in identifying and confirming new
multiple-image systems, thus allowing more detailed modelling.  In
addition, not only the centre of the light for each of the arcs can be
used as constraints, but also their morphology.  As mentioned in
\citetalias{bradac04a}, the reconstruction technique with adaptive
grid at image positions can be used for these purposes. Further,
spectroscopic redshifts need to be obtained for the multiple-image
system candidates as well as for the cluster members (to obtain
velocity dispersion measurements from a large sample). Deep,
wide-field imaging data of this cluster will help us to improve the
weak lensing constraints also at larger radii than presented here. A
large number density of sources that can be used for weak lensing
accessible by ACS ($\gtrsim$ $120\mbox{ arcmin}^{-2}$) would greatly
improve the accuracy of the mass estimate and enable us to resolve
substructures in the cluster. The details of the reconstruction can be
used to reliably determine the cluster profile.

In conclusion, even without the best data quality that can be reached
at present, we were able to perform a detailed cluster-mass
reconstruction of the most X-ray luminous cluster {\RXJ}. The method
has also shown a high potential for the future. If the highest quality
data is used, a combination of strong and weak lensing has proven to
offer a unique tool to pin down the masses of galaxy-clusters as well
as their profiles and accurately test predictions within the CDM
framework.

\begin{acknowledgements}
  We would like to thank L\'{e}on Koopmans, Oliver Czoske, J\"{o}rg
Dietrich, and Thomas Reiprich for many useful discussions that helped
to improve the paper. Further we would like to thank Volker Springel
for providing us with the simulations used in the first paper and
Myriam Gitti for providing us with the X-ray mass estimates. We also
thank our referee for his constructive comments. This work was
supported by the International Max Planck Research School for Radio
and Infrared Astronomy, by the Bonn International Graduate School and
the Graduiertenkolleg GRK 787, by the Deutsche Forschungsgemeinschaft
under the project SCHN 342/3--3, and by the German Ministry for
Science and Education (BMBF) through DESY under the project
05AE2PDA/8. MB acknowledges support from the NSF grant
AST-0206286. This project was partially supported by the Department of
Energy contract DE-AC3-76SF00515 to SLAC.
\end{acknowledgements}
\bibliography{/home/marusa/latex/inputs/bibliogr_clusters,/home/marusa/latex/inputs/bibliogr,/home/marusa/latex/inputs/bibliogr_cv}
\bibliographystyle{/home/marusa/latex/inputs/aa}
\end{document}